# GRAVOTHERMAL OSCILLATIONS

Douglas C. Heggie

University of Edinburgh,
Department of Mathematics and Statistics,
King's Buildings,
Edinburgh EH10 5NL,
U.K.
E-mail: d.c.heggie@ed.ac.uk

**Abstract**

Gravothermal oscillations occur in several models for the post-collapse evolution of rich star clusters. This paper reviews the main literature, and presents some new $N$-body results which appear to exhibit this phenomenon.

## 1. Introduction

In the early 80s Sugimoto and Bettwieser (Sugimoto & Bettwieser 1983, Bettwieser & Sugimoto 1984) were studying the evolution of globular star clusters using a gas model, i.e. the two-body relaxation was treated in analogy with thermal conduction in gases. In addition they included a model for the way in which thermal energy (i.e. the translational kinetic energy of individual stars) was generated in the formation and evolution of binary stars, so that the evolution could be followed past core collapse. They found that the post-collapse evolution was *oscillatory*, the central density varying over several orders of magnitude on a time scale of thousands of elapsed central relaxation times.

The discovery of gravothermal oscillations came as a complete surprise, because until then it had been thought (Heggie 1984, Cohn 1985) that the post-collapse evolution should be steady. Following a suggestion by Sugimoto, it was then found that the steady evolution observed by these authors was caused by use of too large a time-step in the numerical integrations (cf. Heggie & Ramamani 1989, Cohn et al 1989), though steady evolution is the correct behaviour if the strength of the heating is sufficiently high.

The essential ingredients for the occurrence of gravothermal oscillations were already listed in the discovery papers of Bettwieser and Sugimoto. In addition to relaxation and energy generation, which we have already mentioned, a crucial factor is the *gravothermal behaviour* of stellar systems, i.e. phenomena which can be understood in terms of a negative specific heat (Lynden-Bell & Wood 1968, Hachisu & Sugimoto 1978). The best known example of this is the phenomenon of *core collapse*, which arises when the core is slightly hotter than its surroundings. This is also the situation in the collapse phase of gravothermal oscillations. If there is a "temperature inversion", so that the core is slightly cooler than its immediate surroundings, then there is a gravothermal expansion (Hachisu et al 1978), such as occurs in the expansion phase of the oscillations. The switch between collapse and expansion occurs when the generation of energy becomes sufficiently fast (at high enough densities), while the switch between expansion and contraction occurs when the surroundings of the core, where the temperature is highest, come into good thermal contact with the relatively cool halo of the entire system, and the temperature inversion can diffuse away.

In the years since the discovery of the oscillations, considerable detail has been added to the above basic picture. It is one purpose of this paper to review the principal literature on the topic, which has been studied using different models and from several points of view. This is the subject of §2. Our second aim (§3) is to examine in somewhat greater detail the occurrence of gravothermal oscillations in $N$-body simulations, which has been one of the more long-standing controversies within this area. The paper ends with some remarks on future problems which remain to be tackled, especially in the $N$-body context.

## 2. Studies of Gravothermal Oscillations

2.1 A simple model

The mechanisms which are thought to be responsible for the oscillations (§1) can be built into a simple model which describes the evolution of the temperature in three zones: the core, its immediate surroundings, and the halo of the system (Allen & Heggie 1992). With some shortcomings, this model exhibits many of the aspects of gravothermal oscillations which are observed in more elaborate models.

2.2 Gas models

This is the context in which gravothermal oscillations were discovered and the fundamental physical mechanisms elucidated (cf.§1 and reviews by Bettwieser (1985a) and Sugimoto (1985)). The next step (Bettwieser 1985b) was the extension to multicomponent systems, i.e. those in which not all stars have the same mass. In two-component systems it was found that the occurrence of oscillations depended not only on the strength of the energy generation mechanism (as in one-component models) but also on the relative numbers of low- and high-mass stars.

The dependence on the strength of the energy generation was elucidated by Goodman (1987), who showed that this should really be understood as a dependence on the number of stars in the system (assuming that the mechanism of energy generation is the formation and evolution of binaries in three-body encounters). He confirmed that steady post-collapse expansion occurs if the strength is high (corresponding to small $N \lesssim 7000$), while large-amplitude, irregular oscillations occur for large enough systems. At intermediate values of $N$ oscillations of modest amplitude occur, and the transition exhibits features in common with the "period-doubling route to chaos" (e.g. Cvitanović 1989). Goodman's conclusions were concerned with the stability of a certain family of models constructed so as to be expanding initially, but by and large they carry over into models exhibiting expansion after the usual initial period of core collapse (Heggie & Ramamani 1989). Heggie & Aarseth (1992) briefly reported the occurrence of oscillations in the post-collapse evolution of a gaseous system powered by primordial binaries.

All these models assume that the velocity dispersion is locally anisotropic, but recent developments in the formulation of *anisotropic* gas models are now sufficient to allow explorations of gravothermal oscillations in their post-collapse evolution (Spurzem & Louis 1993). The gravothermal properties of anisotropic models are also relevant to this aspect of the problem (Spurzem 1991).

These models also assume that the generation of energy is smooth (proportional to simple powers of the density and temperature), whereas in $N$-body systems the interactions of binaries are essentially stochastic. Inagaki & Hut (1988) considered the effect of stochastic energy sources, within the context of a very simple model for the evolution of the core. M. Giersz has also modified a gas code (Heggie & Ramamani 1989) to include the effects of the stochastic formation and hardening of binaries. This task was undertaken as part of a large comparative study of gas- and $N$-body models (Giersz & Heggie 1993, where further details can be found). Fig.1 shows the evolution of the central density for a model with $N = 5000$. The oscillations are more regular than in a previous study using stochastic processes in a Fokker-Planck model (Takahashi & Inagaki 1991, cf.§2.3), but have comparable depth and duration. The important point is that this model would exhibit steady expansion if the generation of energy were smooth.

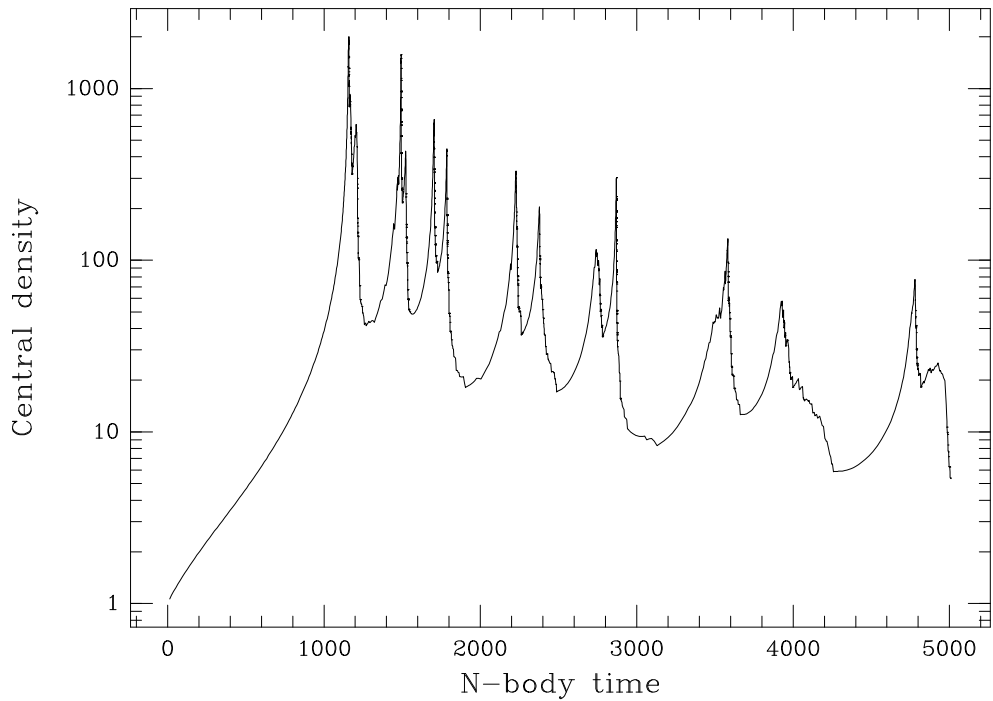

Fig.1 Central density in a gas model with stochastic formation and hardening of binaries. Initial conditions are constructed from Plummer's model, all stars having the same mass, and the system is isolated. Units are standard (Heggie & Mathieu 1986).

2.3 Fokker-Planck models

This is the main tool for the modelling of large star clusters, and the occurrence of gravothermal oscillations in the post-collapse evolution of isotropic, single-component Fokker-Planck models was reported by Cohn *et al* (1986). Actually a second component was included in some of their models, in which binaries were assumed to form by tidal (two-body) interactions (cf. Statler *et al* 1987).

Stochastic generation of energy by three-body binaries was incorporated into a Fokker-Planck treatment by Takahashi & Inagaki (1991), who found that this tends to destabilise small clusters. Clusters which would expand steadily if the generation of energy were smooth were found to exhibit oscillations of large amplitude (e.g. about 1 dex or larger in the central density when $N = 5000$), while clusters of intermediate size, which exhibit small-amplitude oscillations if energy is provided continuously, show irregular oscillations, of larger amplitude, if the source of energy is stochastic. Actually, the first Fokker-Planck models with stochastic input of energy appear to have been those computed (using a Monte Carlo method) by Stodółkiewicz (M. Giersz, pers. comm.), but they remained unpublished at the time of his death. His models included the effects of anisotropy.

Most other work has been carried out with the standard isotropic Fokker-Planck equation, and smooth energy generation. Cohn *et al* (1989) and Breeden *et al* (1992) incorporated three-body binary heating in one-component systems, and the latter authors gave an especially careful account of the transition from stable post-collapse expansion to large-amplitude chaotic oscillations as the number of stars is increased through the appropriate range. Murphy *et al* (1990) demonstrated clearly that inclusion of a reasonably realistic spectrum of masses (between 0.1 and $1.2M_\odot$) suppressed the occurrence of gravothermal oscillations, which arose only in systems whose total mass exceeded roughly $8 \times 10^4 M_\odot$. The steepness of the mass spectrum is also an important factor, and the oscillations affect the different components to very different extents. Gao *et al* (1991) described the occurrence of gravothermal oscillations in a model powered by primordial binaries.

The subject was taken in a new direction by Breeden *et al* (1990), who applied concepts of dynamical systems theory (e.g. Liapounov exponents, first return maps) to explore the transition to chaotic oscillations as $N$ is increased. This links most closely with the model described in §2.1 above.

2.4 Observational models

Though there is considerable evidence for the view that some Galactic globular clusters have passed core collapse (e.g. Djorgovski & King 1986, Lugger *et al* 1987, Aurière & Ortolani 1989, Hut & Djorgovski 1992, Trager *et al* 1993, Drukier 1993), and even some extragalactic ones (Meylan & Djorgovski 1987), the role of gravothermal oscillations is less clear. The main point was made by Bettwieser & Fritze (1984), who remarked that a cluster undergoing gravothermal oscillation is most likely to be observed during the long-lived expansion phase of an oscillation. This remark underlies the application of the Fokker-Planck model of Grabhorn *et al* (1992) to M15, and the interpretation of the nucleus of the galaxy M33 offered by Hernquist *et al* (1991).

## 3. $N$-Body Studies

3.1 Previous work

The ideal tool for discussion of the dynamical evolution of stellar systems is $N$-body simulation, and yet it is here that the occurrence of gravothermal oscillations has proved most controversial and elusive. Makino *et al* (1986) and Makino & Sugimoto (1987) reported oscillations in systems with $N = 100$ and 1000 stars, and concluded that the oscillations were essentially gravothermal, though complicated by small-$N$ effects in the smaller systems. This was partially disputed by Heggie & Ramamani (1989), who found that the oscillations in the 100-body models were taking place on much too short a time scale.

The most forceful objections were made by Inagaki (1986), who not only found no oscillations in a 3000-body system with unequal masses, but gave a general argument for their non-existence in $N$-body systems. Essentially, he argued that the fluctuations in a small core and its surroundings are large enough to mask the temperature inversion which is apparently required (§1) to drive the expansion phase of the oscillations. In the first place, however, the evidence of his $N$-body calculation is weakened by the presence of a mass spectrum, since gas- and Fokker-Planck models show that this tends to suppress the oscillations. Secondly, the later work of Takahashi & Inagaki (1991) shows that the stochastic effects of binary formation and interactions enhance the tendency to gravothermal behaviour, and so it is not clear that fluctuations in the temperature profile will be as effective in smothering the temperature inversion as in models with smooth energy generation.

It should be repeated, however, that, according to Inagaki's original argument, the most serious potential obstacle to the occurrence of gravothermal oscillations arises at the expansion phase. This appears to depend on a temperature inversion of only a few per cent (i.e. the relative difference between the temperature of the core and the maximum value of the temperature, which occurs some distance outside the core radius), which could be masked by Poisson fluctuations in the distribution of velocities, even in systems so large that the relevant region contains thousands of stars. By contrast, no-one doubts that core collapse occurs in $N$-body systems, or that binaries can reverse the collapse, though the direct evidence that a temperature inversion is set up in an $N$-body model is apparently restricted to the results of Makino & Sugimoto (1987).

If the occurrence of gravothermal expansion in an $N$-body system is in doubt, the question can be settled in principle by $N$-body simulations. This was the motivation for a pilot study carried out by Heggie (1989). The first step in this investigation was to select a gas model which exhibited well developed gravothermal oscillations sustained by a simple model for three-body binary heating. Next, a time was selected at which the model was undergoing a gravothermal expansion powered by a temperature inversion. Ideally the next step would have been to create initial conditions for an $N$-body simulation representing the entire system. However, this was impractical, since the gas model corresponded to a system with $N = 20000$. However, it is known from gas and Fokker-Planck models that gravothermal oscillations only involve the innermost few percent of each system, and it was verified that the inner parts of the gas model would undergo a gravothermal expansion followed by recollapse even if the outer parts were replaced by a reflecting spherical wall. Therefore only the innermost parts (about 15% of the system) were modelled by an $N$-body system, and the resulting initial conditions were integrated using a simple $N$-body code with a reflecting boundary. The results were qualitatively consistent with those predicted by the gas model: the evolution of the Lagrangian radii exhibited the same pattern, though the rate of expansion of the core was almost twice as large as in the gas model. The Lagrangian radii also exhibited substantial fluctuations, some of which are now known to be kinematic in origin (Sweatman 1993).

At face value this result supports the view that gravothermal expansion can occur in $N$-body systems even if the temperature inversion is as small as in the gas models. Nevertheless, Heggie's investigation is subject to criticism on two grounds: first, that the run was relatively short, and, second, that other runs might well behave differently, especially if stochastic effects are of possible importance. These criticisms have subsequently been answered (Heggie *et* al 1993), as there are now results from four further independent runs, the longest of them extending the duration of Heggie's original run by a factor of over 20. All the qualitative conclusions reported in Heggie (1989) are confirmed. The main quantitative change is that the expansion of the $N$-body systems is in general *smaller* than that predicted by the gas model, by almost a factor of two. In addition, these long runs yielded considerable information on other interesting, and poorly understood aspects of collisional stellar dynamics, including power spectra of the motion of the core, and of the fluctuations in the Lagrangian radii.

3.2 A new $N$-body result

The main new results of this paper are contained in the present section, which describes results of a 5000-body calculation carried out with the code NBODY5 (Aarseth 1985). Initial conditions were constructed from Plummer's model with stars of equal mass, and the system is isolated. Results are presented in standard units (Heggie & Mathieu 1986), in which the initial crossing time is $2\sqrt{2}$. The calculation to $t = 4437$ took over 2600 hours on a Dec Alpha.

The evolution of the central density (strictly, the mean density within the sphere, centred at the density centre and containing 1% of the bound mass) is shown in Fig.2 The main feature of interest here is the low-density phase in the interval from $t \simeq 2000$ to $t \simeq 2500$. Both in its depth and its duration it resembles the major oscillations present in Fig.1. Indeed the entire run looks quite similar to the Fokker-Planck models, with stochastic energy generation, presented by Takahashi & Inagaki (1991) for the same particle number (their Figs.2(a) and 3(a)).

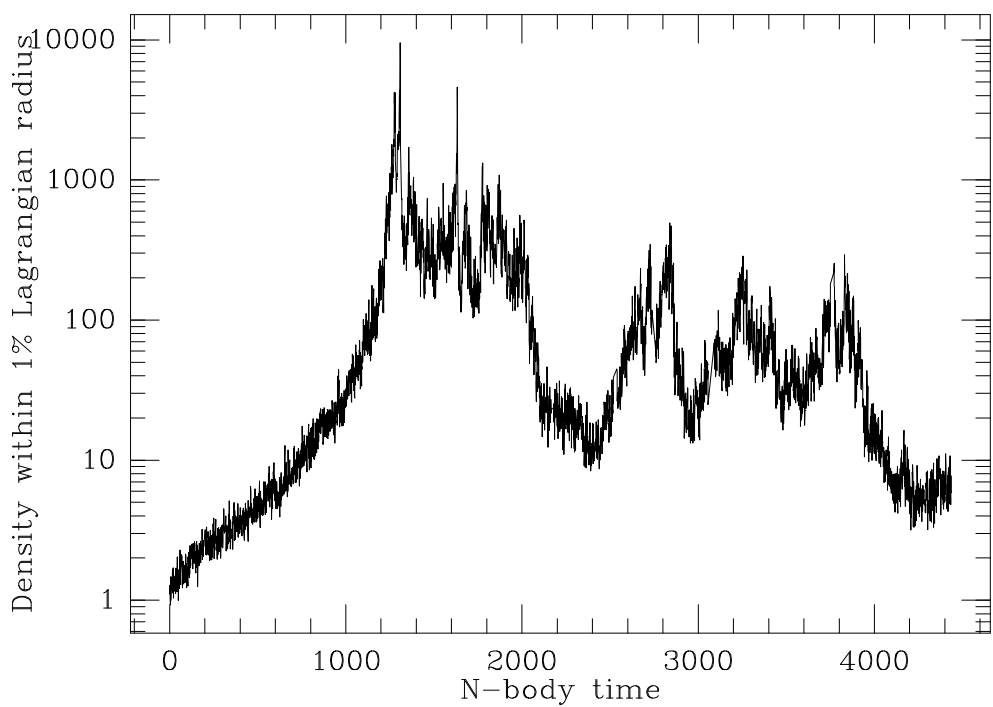

Fig.2 Mean density in the innermost 1% of a 5000-body system. Code, units, and initial and boundary conditions are explained in the text.

This visual resemblance is the main argument for supposing that the oscillation in the $N$-body model in Fig.2 has the same nature as those observed in Fokker-Planck and gas models. Kinematic data (on the mean square speed of stars in several Lagrangian shells) is available, but no trace of a temperature inversion can be found, even if the results are averaged over the duration of the main expansion phase. This argument is not decisive, partly on statistical grounds, and partly because the gas model shown in Fig.1 also exhibits no temperature inversion if the temperature is averaged over each Lagrangian shell, as in the $N$-body model. The absence of a temperature inversion in either model does not imply that the expansion phases in these models are not gravothermal, for the following reason. A system in *steady* post-collapse expansion would exhibit a negative temperature gradient, and the presence of oscillations of sufficiently small amplitude would not create a temperature inversion. On the other hand, it is possible that there exists a temperature inversion in the models, but that it is masked by averaging over fixed Lagrangian shells.

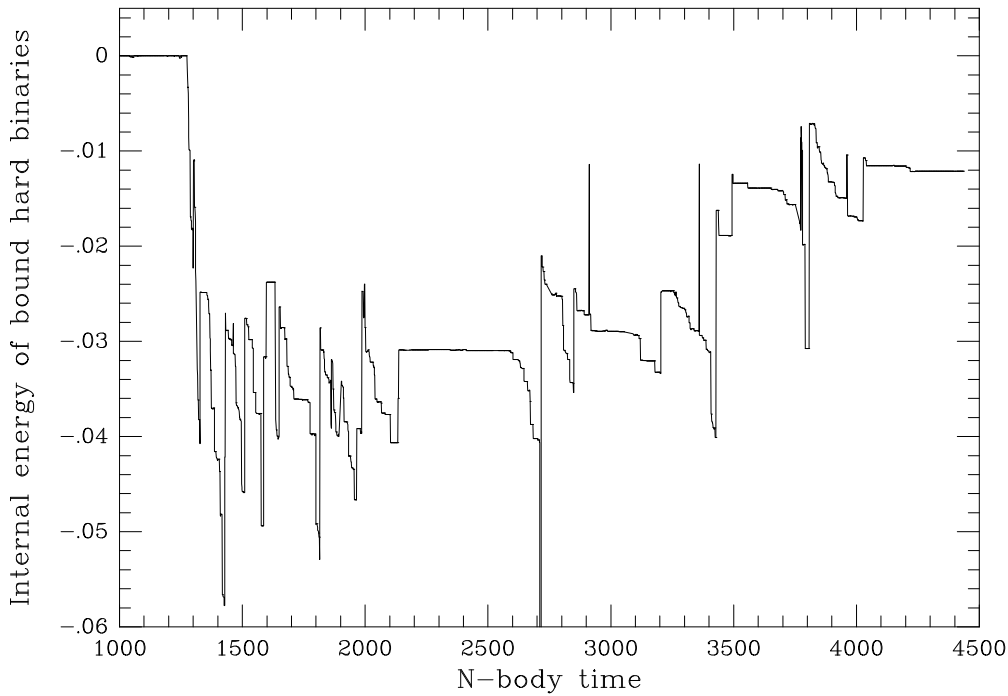

Fig.3 Internal energy of bound hard binaries in the 5000-body system shown in Fig.2. "Internal energy" is the energy of relative motion and mutual potential energy of the components; "hard binaries" are here taken to be defined by those which are regularised in the $N$-body code.

The other possible explanation of the expansion in the $N$-body model is direct heating of the core by binary activity, or indirect heating (caused by the ejection of mass from the core following interactions involving binaries). For this reason Fig.3 shows the internal energy of bound hard binaries. In this curve, a decrease usually corresponds to the hardening of a single hard binary, and then the abrupt increase occurs when its barycentre finally acquires sufficient translational kinetic energy to escape from the system. The escape of a binary just before $t = 2000$ significantly precedes the initiation of the prominent expansion under discussion. A remaining hard binary does harden significantly during the expansion, and then itself escapes at a time significantly later than the end of the expansion phase. Even if this binary is implicated in the expansion, it does not provide a complete explanation of it, partly because of the lack of synchronisation, and partly because there are several other phases of binary hardening and ejection which are not associated with comparable expansions. It is also clear from the duration of the expansion (which lasts almost 80 time units) that it is not the response to any single ejection from the core.

## 4. Conclusions and Discussion

The simplified models for the study of the evolution of large star clusters are unanimous in predicting the occurrence of gravothermal oscillations in the post-collapse phase of the evolution, whether this is powered by binaries formed in three-body encounters, tidal-capture binaries, or primordial binaries, and whether the generation of energy is smooth or stochastic. The most important factor which tends to suppress the oscillations is the presence of a spectrum of stellar masses. Apparently relatively little has been done to investigate the effect of a tidal field. This might not be expected to have a significant influence on a phenomenon which only involves the innermost few percent of the stars, though it is known to have a sizable effect on the rate of core collapse (Spitzer & Chevalier 1973).

Undoubtedly the area where most work is still needed is in the context of $N$-body models. The evidence for core collapse, and core bounce associated with the formation and evolution of binaries in three-body encounters, is quite sound, but the only reasonably persuasive evidence for complete cycles of gravothermal oscillations comes from the simulation studied by Makino & Sugimoto (1987), with $N = 1000$, and the new 5000-body results described in §3.2. These cannot be understood in terms of the gravothermal oscillations exhibited by simple gas and Fokker-Planck models with smooth binary heating, because even the 5000-body system is somewhat too small. One possibility is that these simplified models overestimate the effective rate of heating (as this would tend to suppress oscillations at a given value of $N$). Another possibility is that the stochastic nature of binary heating in the $N$-body model enhances the tendency to gravothermal oscillations, as it does in suitably modified gas and Fokker-Planck models (§§2.2, 2.3). The evidence from the $N$-body models themselves is ambiguous. A binary is active in the most prominent expansion shown by the model discussed in §3.2, but does not appear to be the sole cause of the expansion. Makino & Sugimoto did report a temperature inversion after the core had expanded, but gas models indicate that the temperature inversion should have disappeared by this time.

The work of Heggie (1989) and Heggie *et al* (1993) confirms one of the mechanisms which is thought to be fundamental to the occurrence of the oscillations (viz, gravothermal expansion driven by a small temperature inversion), but did not establish that a suitable temperature inversion would be set up, and took some liberties with the boundary conditions.

Advances in $N$-body technique, especially the development of GRAPE and HARP, should soon make it possible to establish unambiguously whether fully developed gravothermal oscillations can occur in $N$-body systems. For this purpose the oscillations should be sufficiently large in amplitude. Studies of the gas model (Heggie & Ramamani 1989) and the Fokker-Planck model (Breeden *et al* 1992) show that the amplitude of the oscillations, as measured by the central density, increases from about 1 dex at $N = 10000$ to about 3 dex when $N = 20000$. On the other hand when $N = 50000$ the amplitude has increased only to 3.5 dex, roughly. This suggests that an $N$-body simulation with $N \sim 20000$ offers a reasonable and practical compromise between the effort required and the ease of detection of the oscillations. Proving that the oscillations are really gravothermal is more difficult. In the experiments described by Heggie (1989) and Heggie *et al* (1993) a temperature inversion was built into the initial conditions, and, in the absence of significant binary activity, was the only plausible explanation of the behaviour observed. In future $N$-body explorations some care will be required to measure the temperature inversion expected (cf. Makino & Sugimoto 1987), and to establish that the behaviour is not simply driven by the evolution of binaries. One way in which this could be done is to identify the start of a gravothermal expansion, remove or sterilise any active hard binary, and recompute the evolution to see whether the expansion persists.

## Acknowledgements

A previous version of this paper was presented at a workshop on "Astrophysics with Special-Purpose Computers", held at the University of Tokyo on August 23 and 24, 1993. The author's participation was funded through a grant from the Japanese Ministry of Education, Science, and Culture, and he is indebted to Prof. D. Sugimoto and other members of the GRAPE team for their kindness and hospitality on that occasion. Thanks are due to B.R.P. Murdoch for permission to use the Alpha operated by Edinburgh University Computing Service, to S.J. Aarseth for the provision of NBODY5, to M. Giersz and J.A. Blair-Fish for preparing this code for the Alpha, and to M. Giersz for permission to show Fig.1. Dr Giersz's research at Edinburgh was supported by the U.K. Science and Engineering Research Council under grant No. GR/G04820.